\documentclass[prd,twocolumn,aps,amsmath,amssymb,nofootinbib,preprintnumbers]
{revtex4}  \voffset=1cm
\usepackage{graphicx}
\usepackage{dcolumn}
\usepackage{bm}
\usepackage{amsmath}
\usepackage{amsfonts}

\def\ls{\mathrel{\lower4pt\vbox{\lineskip=0pt\baselineskip=0pt
           \hbox{$<$}\hbox{$\sim$}}}}
\def\gs{\mathrel{\lower4pt\vbox{\lineskip=0pt\baselineskip=0pt
           \hbox{$>$}\hbox{$\sim$}}}}
\def\drawbox#1#2{\hrule height#2pt

\hbox{\vrule width#2pt height#1pt \kern#1pt
              \vrule width#2pt}
              \hrule height#2pt}

\def\Asym#1#2{\vcenter{\vbox{\drawbox{#1}{#2}
              \kern-#2pt       
              \drawbox{#1}{#2}}}}

\newcommand{\beq}{\begin{equation}}
\newcommand{\eeq}{\end{equation}}

\begin{document}
\preprint{MIFP-07-20}
\title{Unifying inflation and dark matter with neutrino masses}

\author{Rouzbeh Allahverdi$^{1,2}$}
\author{Bhaskar Dutta$^{3}$}
\author{Anupam Mazumdar$^{4,5}$}

\affiliation{$^{1}$~Perimeter Institute for Theoretical Physics, Waterloo, ON, N2L 2Y5, Canada \\
$^{2}$~Department of Physics and Astronomy, University of New
Mexico, Albuquerque, NM 87131, USA \\
$^{3}$~Department of Physics, Texas A\&M University, College Station, TX 77843-4242, USA\\
$^{4}$~NORDITA \& Niels Bohr Institute, Blegdamsvej-17, Copenhagen-2100, Denmark\\
$^{5}$~Physics Department, Lancaster University, LA1 4YB, UK}

\begin{abstract}
We propose a simple model where a {\it gauge invariant inflaton} is
responsible for cosmic inflation and generates the seed for
structure formation, while its relic {\it thermal} abundance
explains the missing matter of the universe in the form of cold dark
matter. The inflaton self-coupling
also explains the observed neutrino masses. All the virtues can be
attained in a minimal extension of the Standard Model gauge group
around the TeV scale. We can also unveil these properties of an
inflaton in forthcoming space and ground based experiments.
\end{abstract}
\maketitle


There are three important puzzles, the origin of inflation, the
origin of cold dark matter, and the origin of neutrino masses, which
require explanation in any extension of the electroweak standard
model (SM). The aim of this paper is to bind the three issues
together and explain all of them in a single set up.
Inflation is driven by a scalar field known as the {\it inflaton},
for a review see~\cite{AM-REV}. In this paper we shall illustrate an
inflaton part of which directly decays into the SM baryons, its {\it
thermal} relic abundance accounts for the missing matter in the
universe, commonly known as cold dark matter, and also explains the
observed neutrino masses.

The inflaton potential explains the flatness of the universe and
also the observed anisotropy in the cosmic microwave background
(CMB) temperature, i.e. $\delta_{H}\sim 1.91\times 10^{-5}$ with an
observed range of spectral index: $0.92 \leq n_s\leq 1.0$ (within
$2\sigma$ error bar)~\cite{WMAP3}. The inflaton carries SM gauge
charges, and hence naturally produces (a thermal bath of) the SM
degrees of freedom~\cite{AEGM,AKM}. If the reheat temperature of the
universe is higher than the mass of the inflaton, then the plasma
upon reheating will, in addition, have a thermal distribution of
the inflaton quanta. If the inflaton is absolutely stable, due to
some symmetry, then it can also serve as
the cold dark matter. Once the temperature drops below the inflaton
mass, its quanta undergo thermal freeze-out and may yield the
correct dark matter abundance. As a consequence, direct and indirect
dark matter detection experiments will provide valuable information
about the inflaton couplings to the SM fields.

The main question is whether relic density of the inflaton after
thermal freeze-out can account for dark matter in the universe. For
particles with gauge interactions, the unitarity bound puts an
absolute upper bound $\sim 100$ TeV on their mass~\cite{GK}. For
smaller couplings this bound will decrease in order to have
acceptable thermal dark matter, for a review see~\cite{CDM-REV}.
Therefore, the inflaton mass should be below $100$~TeV, which puts a
severe constraint on the flatness of the inflaton potential. As we
will show an inflaton with a weak scale mass and a tiny
self-coupling can drive a successful inflation.
Coincidentally such a small coupling is welcoming if the neutrinos
are Dirac in nature~\cite{NEUT-REV}. It turns out that if the
inflaton is composed of the SM Higgs, slepton and the sneutrino,
then it serves all three purposes: inflation, cold dark matter, and
neutrino mass.

Let us consider the minimal supersymmetric standard model (MSSM)
with three right-handed (RH) neutrino multiplets. We assume the
neutrinos to be of Dirac type. Whether the nature of neutrino is
Dirac or Majorana can be determined in the future neutrinoless
double beta decay experiment. Then relevant superpotential term is

\beq \label{supot} W  \supset h {\bf N} {\bf H}_u {\bf L} . \eeq
Here ${\bf N}$, ${\bf L}$ and ${\bf H}_u$ are superfields containing
the RH neutrinos, left-handed (LH) leptons and the Higgs which gives
mass to the up-type quarks, respectively. For conciseness we have
omitted the generation indices, and we work in a basis where
neutrino Yukawa couplings $h$ (and hence neutrino masses) are
diagonalized.

For various reasons, which will become clear, the inclusion of a
gauge symmetry under which the RH (s)neutrinos are not singlet is
crucial. As far as inflation is concerned, a singlet RH sneutrino
would not form a gauge-invariant inflaton along with the Higgs and
slepton fields. We consider a minimal extension of the SM gauge
group which includes a $U(1)$: $SU(3)_c \times SU(2)_L \times U(1)_Y
\times U(1)_{B-L}$, where $B$ and $L$ denote the baryon and lepton
numbers respectively. This is the simplest extension of the SM
symmetry which is also well motivated: anomaly cancelation requires
that three RH neutrinos exist, so that they pair with LH neutrinos
to form three Dirac fermions.

In this model we have an extra $Z$ boson ($Z^{\prime}$) and one
extra gaugino ($\tilde Z'$). The $U(1)_{B - L}$ gets broken around
TeV by new Higgs fields with ${B - L} = \pm 1$.  This also prohibits
a Majorana mass for the RH neutrinos at the renormalizable level
(note that ${\bf N N}$ has ${B - L} = 2$). The Majorana mass can be
induced by a non-renormalizable operator, but it will be very small.

The value of $h$ needs to be small, i.e. $h \ls 10^{-12}$, in order
to explain the light neutrino mass, $\sim {\cal O}(0.1~{\rm eV})$
corresponding to the atmospheric neutrino oscillations detected by
Super-Kamiokande experiment~\footnote{One can assume that the
neutrino mass terms are part of the K\"ahler potential
Ref.~\cite{Arnowitt:2003kc}. One can then generate a small neutrino
mass in the superpotential by K\"ahler transformation (similar to
generation of $\mu$ parameter).}.

Note that the ${\bf N} {\bf H}_u {\bf L}$ monomial represents a
$D$-flat direction under the $U(1)_{B-L}$, as well as the SM gauge
group~\cite{MSSM-REV}.
(a flat direction represented by a polynomial is more
involved~\cite{JOKINEN}). The flat direction field $\phi$ is defined
as
\beq \label{flat} {\phi} = {{\tilde N} + {H}_u + {\tilde L} \over
\sqrt{3}} , \eeq
where ${\tilde N}$, ${\tilde L}$, $H_u$ are the scalar components of
corresponding superfields. Since the RH sneutrino ${\tilde N}$ is a
singlet under the SM gauge group, its mass receives the smallest
contribution from quantum corrections due to SM gauge interactions,
and hence it can be set to be the lightest supersymmetric particle
(LSP). Therefore the dark matter candidate arises from the RH
sneutrino component of the inflaton, see Eq.~(\ref{flat}). The
potential along the flat direction, after the minimization along the
angular direction, is found to be~\cite{AKM},
\begin{eqnarray} \label{flatpot}
V (\vert \phi \vert) = \frac{m^2_{\phi}}{2} \vert \phi \vert ^2 +
\frac{h^2}{12} \vert \phi \vert^4 \, - \frac{A h}{6\sqrt{3}}
\vert \phi \vert^3 \,.
\end{eqnarray}
The importance of $A$-term in providing large vacuum energy was
shown before~\cite{AEJM}.  The flat direction mass $m_{\phi}$ is
given in terms of ${\tilde N},~H_u,~{\tilde L}$ masses:
\beq \label{phimass} m^2_{\phi} = {m^2_{\tilde N} + m^2_{H_u} +
m^2_{\tilde L} \over 3}. \eeq
Note that for $A = 4 m_{\phi}$, there exists a {\it saddle point} for
which $V^{\prime}(\phi_0) = V^{\prime \prime}(\phi_0) = 0$. The saddle
point and the potential are given by:
\begin{eqnarray} \label{sad} \phi_0 = \sqrt{3}\frac{m_{\phi}}{h}=
6 \times 10^{12} ~ m_{\phi} ~ \Big({0.05 ~
{\rm eV} \over m_\nu} \Big)\,, \\
\label{sadpot}
V(\phi_0) = \frac{m_{\phi}^4}{4h^2}=3 \times 10^{24} ~ m^4_{\phi} ~
\Big({0.05 ~ {\rm eV} \over m_\nu} \Big)^2 \,.
\end{eqnarray}
Here $m_\nu$ denotes the neutrino mass which is given by $m_\nu = h
\langle H_u \rangle$, with $\langle H_u \rangle \simeq 174$ GeV. For
neutrino masses with a hierarchical pattern, the largest neutrino
mass is $m_\nu \simeq 0.05$ eV in order to explain the atmospheric
neutrino oscillations~\cite{atmos}, while the current upper bound on
the sum of the neutrino masses from cosmology, using WMAP and SDSS
data alone, is $0.94$ eV~\cite{wmapsdss}.

Inflation can take place along the {\it gauge invariant} flat
direction $\phi$, near a saddle point $\phi_0$~\footnote{This
requires that the inflaton starts in the vicinity of $\phi_0$ with
$\dot\phi_0\approx 0$~\cite{AFM}. Around a saddle point there exists
a self-reproduction regime such that most parts of the universe
inflate forever~\cite{AKM}.}. The dynamics of the inflaton can be
understood by expanding the potential,
$V(\phi)=V(\phi_0)+(1/3!)V^{\prime\prime\prime}(\phi_0)(\phi-\phi_0)^3+...$,
see Refs.~\cite{AKM,AEGM}. The slow roll inflation is governed by
the third derivative of the potential,
$V^{\prime\prime\prime}(\phi_0)=(2/\sqrt{3})hm_{\phi}$. A sufficient
number of e-foldings is generated during the slow roll, i.e. ${\cal
N}_e\sim (\phi_0^2/m_{\phi}M_{\rm P})^{1/2}$~\cite{AKM}. For
$m_{\phi}\sim 50$~GeV, it turns out to be ${\cal N}_e \sim 10^3$.

The amplitude of density perturbations follows~\cite{AKM}.
\beq \label{amp} \delta_{H} \simeq
\frac{1}{5\pi}\frac{H^2_{inf}}{\dot\phi} \simeq 3.5 \times 10^{-27}
~ \Big( {m_\nu \over 0.05 ~ {\rm eV}} \Big)^2 ~  \Big({M_{\rm P}
\over m_{\phi}} \Big) ~ {\cal N}_{\rm COBE}^2\,. \eeq
Here $m_{\phi}$ denotes the loop-corrected value of the inflaton
mass at the scale $\phi_0$ in
Eqs.~(\ref{sad},\ref{sadpot},\ref{amp})~\footnote{Even though the
inflaton potential gets loop corrections due to gauge interactions,
the existence of a saddle point remains and the flatness of the
potential is not ruined~\cite{AKM}.}. In Figure 1, we show the
neutrino mass as a function of the inflaton mass. We use $\delta_H =
1.91 \times 10^{-5}$ to draw the plot. We see that the neutrino mass
in the range $0$ to $0.30$~eV corresponds to the inflaton mass of
$0$ to $30$~ GeV. We will be using this mass range of the inflaton
to calculate the SUSY masses which allow a RH sneutrino dark matter.

\begin{figure}[t]
\includegraphics[width=7cm]{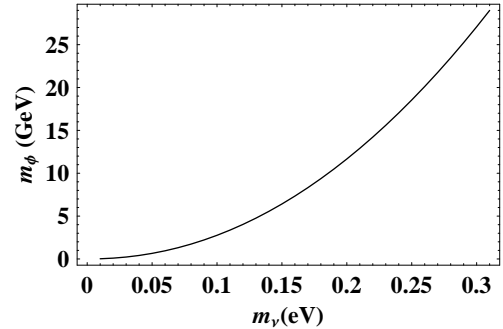}
\caption{The inflaton mass $m_{\phi}$ is plotted as a function of
the neutrino mass $m_{\nu}$.} \label{neutrinoinf}
\end{figure}

The spectral index of the power spectrum is given by $n_s = 1 +
2\eta - 6\epsilon \simeq 1 - {4/ {\cal N}_{\rm COBE}}$, and the
running in the spectral tilt is ${d\,n_s/ d\ln k} = - {4/ {\cal
N}_{\rm COBE}^2} $~\cite{AKM}. Here ${\cal N}_{\rm COBE}$ is the
number of e-foldings between the time the observationally relevant
perturbations were generated and the end of iflation. The exact
number depends on the scale of inflation and on when the Universe
becomes radiation dominated. As we shall see that the universe
becomes radiation dominated instantly (time scale comparable to the
Hubble scale) after the end of inflation. The required number of
e-foldings is rather small, i.e. ${\cal N}_{\rm COBE}\sim
40$~\cite{LEACH}. For such a low ${\cal N}_{\rm COBE}$ the value of
$n_s\sim 0.90$ near the saddle point. At a face value this is
already ruled out by the WMAP data~\cite{WMAP3}. However a slight
departure from a saddle point can provide the spectrum which is well
inside the observed limit. In fact the spectral tilt can be obtained
within $0.90 \leq n_s\leq 1.0$ if $\phi_0$ becomes a point of
inflection, i.e. $V^{\prime}(\phi_0) \neq 0,~V^{\prime
\prime}(\phi_0) = 0$
~\cite{LYTH1,AEGJM}. Note that its
reflection on the amplitude of the density perturbations is
negligible.

The inflaton has gauge couplings to the electroweak and $U(1)_{\rm
B-L}$ gauge/gaugino fields. It therefore induces a VEV-dependent
mass $\sim g \langle \phi \rangle$ for these fields ($g$ denotes a
typical gauge coupling). After the end of inflation, $\phi$ starts
oscillating around the global minimum at the origin with a frequency
$m_{\phi} \sim 10^3 H_{\rm inf}$.  When the inflaton passes through
the minimum, $\langle \phi \rangle = 0$, the induced mass undergoes
non-adiabatic time variation. This results in non-perturbative
particle production~\cite{PREHEAT}. As the inflaton VEV is rolling
back to its maximum value $\phi_0$, the mass of the gauge/gaugino
quanta increases again. Because of their large couplings they
quickly decay to the fields which are not coupled to the inflaton,
hence massless, notably the down-type (s)quarks. This is a very
efficient process as a result of which relativistic particles are
created within few Hubble times after the end of inflation (for more
details, see~\cite{AEGJM}). A thermal bath of MSSM particles is
eventually formed with a temperature $T_{\rm R} \sim 10^6$ GeV (for
details of thermalization in SUSY, see~\cite{AVERDI1}). Note that
the reheat temperature is high enough for the electroweak
baryogenesis~\cite{BARYO-REV}. On the other hand, it is sufficiently
low to avoid overproduction of dangerous relics such as
gravitinos~\cite{AVERDI1,MAROTO}.

Scatterings via the new $U(1)$ gauge interactions also bring the RH
sneutrino into thermal equilibrium. Note that part of the inflaton,
i.e. its ${\tilde N}$ component see Eq.~(\ref{flat}), has never
decayed; only the coherence in the original condensate that drives
inflation is lost.
Its relic abundance will be then set by thermal freeze-out. The fact
that ${\tilde N}$ has gauge interactions under the new $U(1)$ is
crucial in this respect: the neutrino Yukawa $h$ is way too small to
allow acceptable thermal dark matter~\footnote{Or acceptable
sneutrino dark matter at all. Note that ${\tilde N}$ would dominate
the universe right after the end of inflation if it had no gauge
interactions. $h$ is too small to deplete the energy in the ${\tilde
N}$ component of the inflaton via non-perturbative particle
production.}.

Note that the mass of dark matter (i.e. RH sneutrino) is correlated
with the inflaton mass, see Eq.~(\ref{phimass}). However, the former
is calculated at the weak scale, while the latter is at the scale
$\phi_0$, see Eq.~(\ref{sad}). The two quantities are related to
each other by RGEs.  In order to calculate the masses in the model,
we use SUGRA boundary conditions (i.e. $m_0$: universal scalar mass
for the squarks and sleptons (but the Higgs masses are different),
$m_{1/2}$: universal gaugino mass, $A_0$: trilinear scalar coupling,
$\mu>0$ and $\tan\beta$) for the sparticle masses in the extended
$U(1)$ model. We also assume that the gauge couplings are unified at
the GUT scale. Then the new gauge coupling is of the order of
hypercharge gauge coupling.

Even though the inflaton mass is small at the scale $\phi_0$,
obtained by solving Eq.~(\ref{sad}), the RGE effects increase the RH
sneutrino and slepton masses at the weak scale. Note that $m_{\tilde
N}$ and $m_{\tilde L}$ are soft breaking masses, while $m_{H_u}$
also includes the contribution from the $\mu$ term. In Figure 2 we
plot the RH sneutrino and stop masses at the weak scale for
different values of gluino masses. The gluino masses for the three
lines (solid and dashed) are 730 GeV, 1.20 TeV and 1.64 TeV
(bottom-up). The lines are drawn for 0.30 eV neutrino mass. If we
choose the neutrino masses to be 0.05 eV, the lines do not shift
much. The left and right end of the lines correspond to $m_0$=0 and
$m_{\tilde\nu}
> m_{\tilde \chi^0}$. The charged sleptons and the LH sneutrinos are
more massive compared to the RH sneutrino since these particles get
contributions from more gaugino loops due to their $SU(2)_L$ and
$U(1)_Y$ couplings. Figure 2 shows that the masses of sparticles
(e.g., the lighter stop) can be within the reach of initial phase of
the LHC.

\begin{figure}[t]
\includegraphics[width=7cm]{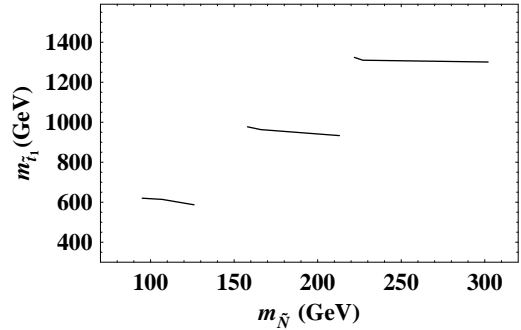}
\caption{$m_{\tilde t_1}$ vs $m_{\tilde N}$. The lines are for
neutrino masses 0.3 eV. The gluino masses for the three lines are
730, 1200 and 1640 GeV (bottom-up). For smaller neutrino masses, the
lines are slightly shifted downwards.} \label{sneutrinoinf}
\end{figure}

In order to calculate the relic abundance of the RH sneutrino, we
need to know the masses of the additional gauge boson $Z^{\prime}$
and its SUSY partner ${\tilde Z}^{\prime}$, the new Higgsino masses,
Higgs VEVs which break the new $U(1)$ gauge symmetry, the RH
sneutrino mass, the new gauge coupling, and the charge assignments
for the additional $U(1)$. We assume that the new gauge symmetry is
broken around 2 TeV~\footnote{This is sufficient to ensure that the
RH neutrinos decouple from the thermal bath early enough in order
not to affect big bang nucleosynthesis (BBN).}, and the existence of
two new Higgs superfields to maintain the theory anomaly free.
The
primary diagrams responsible to provide the right amount of relic
density are mediated by ${\tilde Z}^{\prime}$ in the $t$-channel
. In Figure 3, we show the relic density values for smaller masses
of sneutrino where the lighter stop mass is $ \ls 1$ TeV. The
smaller stop mass will be easily accessible at the LHC and is also
preferred by the little hierarchy solutions~\cite{lh}. We have
varied new gaugino and Higgsino masses and the ratio of the VEVs of
new Higgs fields to generate Fig. 3. We find that the
WMAP~\cite{WMAP3} allowed values of the relic density, i.e.,
$0.094-0.129$ is satisfied for many points. In the case of a larger
sneutrino mass in this model, the correct dark matter abundance can
be obtained by annihilation via $Z^{\prime}$ pole~\cite{matchev}.

\begin{figure}[t]
\includegraphics[width=7cm]{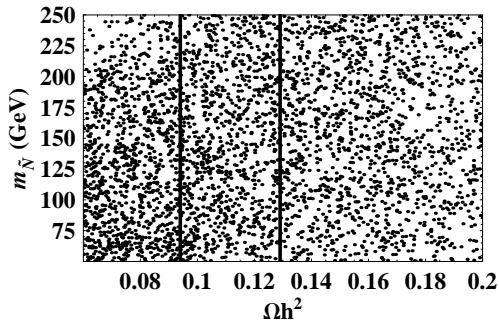}
\caption{$\Omega h^2$ vs $m_{\tilde N}$. The solid lines from left
to right are for $\Omega h^2 =$ 0.094 and 0.129 respectively. The
$Z^{\prime}$-ino mass is equal to the Bino mass since the new $U(1)$
gauge coupling is the same as the hypercharge gauge coupling. 
}\label{sneutrinoominf}
\end{figure}

Since the dark matter candidate, the RH sneutrino, interacts with
quarks via the $Z^{\prime}$ boson, it is possible to see it via the
direct detection experiments. The detection cross sections are not
small as the interaction diagram involves $Z^{\prime}$ in the
$t$-channel. The typical cross section is about 2$\times 10^{-8}$ pb
for a $Z^{\prime}$ mass around 2 TeV. It is possible to probe this
model in the upcoming dark matter detection experiments~\cite{dd}.
The signal for this new scenario at the LHC will contain standard
jets plus missing energy and jets plus leptons plus missing energy.
The jets and the leptons will be produced from the cascade decays of
squarks and gluinos into the final state containing the sneutrino.

In conclusion, within a minimal extension of the SM, we can explain
our universe with observed temperature anisotropy in the CMB, the
right abundance of thermal dark matter, and the Dirac neutrino
masses.
More importantly, this is a scenario which we can test in
laboratory. The dark matter candidate, which is part of the
inflaton, is detectable in direct detection experiments and in the
colliders. Thus it provides us hints on the inflaton, which is
believed to be the most elusive particle in the universe. The future
neutrinoless double beta decay experiments
will lead to valuable information on the nature of neutrinos and
this scenario
.

We wish to thank Alex Kusenko for helpful discussions. The work of
R.A. is supported by Perimeter Institute for Theoretical Physics.
Research at Perimeter Institute is supported in part by the
Government of Canada through NSERC and by the province of Ontario
through MRI. The work of B.D. is supported by the U.S. DOE grant
DE-FG02-95ER40917. A.M. is partly supported by the UNIVERSE-NET
(MRTN-CT-2006-035863).


\end{document}